\magnification=\magstep1
\hsize 6.0 true in
\vsize 9.0 true in
\voffset=-.5truein
\pretolerance=10000
\baselineskip=18truept

\font\tentworm=cmr10 scaled \magstep2
\font\tentwobf=cmbx10 scaled \magstep2

\font\tenonerm=cmr10 scaled \magstep1 
\font\tenonebf=cmbx10 scaled \magstep1

\font\eightrm=cmr8
\font\eightit=cmti8
\font\eightbf=cmbx8
\font\eightsl=cmsl8
\font\sevensy=cmsy7
\font\sevenm=cmmi7

\font\twelverm=cmr12  
\font\twelvebf=cmbx12
\def\subsection #1\par{\noindent {\bf #1} \noindent \rm}

\def\mid {\let\rm=\tenonerm \let\bf=\tenonebf \rm \bf}

\def\para{\par \vskip 12 pt}
 
\def\head{\let\rm=\tentworm \let\bf=\tentwobf \rm \bf}

\def\heading #1 #2\par{\centerline {\head #1} \smallskip
 \centerline {\head #2} \vskip .15 pt \rm}

\def\eight{\let\rm=\eightrm \let\it=\eightit \let\bf=\eightbf 
\let\sl=\eightsl \let\sy=\sevensy \let\m=\sevenm \rm}

\def\foots{\noindent \eight \baselineskip=10 true pt \noindent \rm}
\def\sexion{\let\rm=\twelverm \let\bf=\twelvebf \rm \bf}

\def\section #1 #2\par{\vskip 20 pt \noindent {\mid #1} \enspace {\mid #2} 
  \para \noindent \rm}

\def\abstract#1\par{\para \foots {\bf Abstract: \enspace}#1 \para}

\def\author#1\par{\centerline {#1} \vskip 0.1 true in \rm}

\def\abstract#1\par{\noindent {\bf Abstract: }#1 \vskip 0.5 true in \rm}

\def\sqr#1#2{{\vcenter{\vbox{\hrule height.#2pt
  \hbox {\vrule width.#2pt height#1pt \kern#1pt
  \vrule width.#2pt}
  \hrule height.#2pt}}}}

\def\n{\noindent}
\def\s{\smallskip}
\def\m{\medskip}
\def\b{\bigskip}
\def\c{\centerline}

\def\gne #1 #2{\ \vphantom{S}^{\raise-0.5pt\hbox{$\scriptstyle #1$}}_
{\raise0.5pt \hbox{$\scriptstyle #2$}}}

\def\ooo #1 #2{\vphantom{S}^{\raise-0.5pt\hbox{$\scriptstyle #1$}}_
{\raise0.5pt \hbox{$\scriptstyle #2$}}}


\voffset=-.5truein
\vsize=9truein
\baselineskip=22pt
\hsize=6.0truein
\pageno=1
\pretolerance=10000
.tfm scaled800
\def\n{\noindent}
\def\s{\smallskip}
\def\b{\bigskip}
\def\m{\medskip}
\def\c{\centerline}

\c{\bf\mid Spherically symmetric empty space and its dual in general 
relativity} \b
\b
\b
\c{\bf Naresh Dadhich\footnote{$^*$}{Email :
nkd@iucaa.ernet.in}}
\c{\bf  Inter University  Centre for Astronomy \& Astrophysics}
\c{\bf P.O. Box 4, Pune-411007, India}
\b
\b

 (to appear in Current Science)

{\it In the spirit of the Newtonian theory, we characterize 
spherically symmetric empty space in 
general relativity in terms of energy density measured by a static 
observer and convergence density experienced by null and  timelike 
congruences. It turns out that 
space surrounding a static particle is entirely specified by vanishing 
of energy and null convergence density. The electrograv-dual$^{1}$ to this 
condition would 
be vanishing of timelike and null convergence density which gives the 
dual-vacuum solution representing a Schwarzschild black hole with global 
monopole charge$^{2}$ or with cloud of 
string dust$^{3}$. Here the duality$^{1}$ is defined by interchange of active 
and passive electric parts of the Riemann curvature, which amounts to 
interchange of the Ricci and Einstein tensors. This effective 
characterization of stationary vacuum works for the Schwarzschild and NUT 
solutions. The most remarkable 
feature of the effective characterization of empty space is that it leads 
to new dual spaces and the method can also be applied to lower and higher 
dimensions.}

\n PACS no. 04.00, 04.20 Dw, 04.20 Jb, 98,80 Cq

\vfill\eject

 The Newtonian gravitational field equation is given by 
$\bigtriangledown^2 \phi = 4\pi G\rho$ and empty space is characterized by 
$\rho = 0$.  It is well-known that measure of energy 
is an ambiguous issue in GR primarily because of the inherent difficulty of 
non-localizability of gravitational field energy. However there is no 
difficulty in defining various kinds of energy density, signifying  
different aspects. The analogue of the Newtonian 
matter density is the energy density measured by a static observer and 
defined by $\rho = T_{ab}u^au^b, u^au_a = 1$, where $u_a = 
(\sqrt g_{00},0,0,0)$ and $T_{ab}$ is the 
matter-stress tensor of non-gravitational matter field. Then there is the 
convergence density experienced by timelike and null  
particle congruences in 
the Raychaudhuri equation$^{4}$. They 
are defined as the timelike convergence density, 
$\rho_t = (T_{ab} - {1 \over 2} Tg_{ab})u^a u^b$ and the null convergence 
density $\rho_n = T_{ab}v^av^b, v^av_a = 0, 
v_a = (1,\sqrt{-g_{11}/g_{00}},0,0)$. 
The energy density $\rho$ refers to all kinds of energy 
other than the gravitational field energy, 
while the timelike and null convergence densities act as active 
gravitational charge densities. For perfect fluid they are given by 
$\rho_t = {1 \over 2}(\rho + 3p)$ and $\rho_n = \rho + p$. It is important to 
recognise that 
these three represent different aspects of energy distribution and its 
gravitational linkage. They would thus in general be not equal. Obviously 
all the three can never be equal unless space is flat. However $\rho = 
\rho_t$ implies vanishing of scalar curvature (radiation), $\rho 
= \rho_n$ indicates vanishing of pressure (dust) and $\rho_t = 
\rho_n$ gives $\rho = p$ (stiff fluid). It may be noted that the weak 
field and slow motion limit of the Einstein non-empty space equation is 
$\bigtriangledown^2 \phi = 4\pi G \rho$, while its 
limit in weak field and relativistic motion is 
$\bigtriangledown^2 \phi = 8\pi G \rho_t$.

 In the following, we shall always refer $\rho$ and $\rho_t$ 
relative to a static observer, and $\rho_n$ to radial null 
geodesic. This does however bring in a particular choice for the timelike 
and null vectors but the choice is well motivated by the physics of the 
situation. The radial direction is picked up by the 4-acceleration of the 
timelike particle, identifying the direction of gravitational force, and 
so is the static observer for measure of energy and timelike 
convergence densities...
 
 The main question we wish to address in this note is, can we characterize 
empty space solely in terms of these densities? 

 The answer is yes for the space surrounding a static particle. This may 
in general be true  for isolated particle with some additional 
conditions which would specify the additional physical character of the 
problem. It is clear 
that any specification of empty space must involve density relative to 
both timelike and null particles. That means $\rho_n$ must vanish in 
any case and in addition  one or both of $\rho$ and $\rho_t$ must 
vanish. Of course there should be no energy flux, $P^c = h^{ac} T_{ab}u^b = 
0, h^{ac} = g^{ac} - u^au^c$. It turns out that for spherical symmetry 
effective equation for vacuum is $\rho = \rho_n = P_c = 0$, the solution 
of which would imply $\rho_t = 0$ and vanishing of the all Ricci 
components. Thus 
vanishing of energy and its flux, and null convergence density is 
sufficient to characterize 
empty space for spherical symmetry as these conditions completely 
determine the unique Schwarzschild solution. The effective vacuum 
equation is less restrictive than the vanishing of the entire Ricci tensor. 

 What actually happens is, for the spherically symmetric metric in the 
curvature coordinates, $P_c = 0$ and $\rho_n = 0$ lead to $R_{01} = 
0$ and $R^0_0 = R^1_1$ which imply $g_{00} = f(r) = - g^{11}$, and then 
$\rho = 0$ means $R^2_2 = 0$ which integrates to give the Schwarzschild 
solution completely with $g_{00} = 1 - 2GM/r$ (we have set $c = 1$). Thus 
instead of $R_{ab} = 0$, the less 
restrictive effective equation $\rho = \rho_n = P_c = 0$ also equivalently 
characterizes empty space for a static particle. It 
is a covariant statement relative to a static observer and in 
the curvature coordinates it takes the form $R^0_0 = R^1_1, R^2_2 = 0 = 
R^0_1$.

 Since there are three kinds of density, which could vanish with two at 
a time in three different ways, it is then natural to ask what 
would the other two cases give rise to?

 The first thing that comes to mind is to replace $\rho$ by $\rho_t$ in the 
effective equation to write $\rho_t = 0 = \rho_n = P_c$, which would imply 
$G^0_0 = G^1_1, G^2_2 = 0 = G^0_1$. That is replacing Ricci by Einstein, 
which represents a duality relation between the two. Remarkably 
this duality transformation is implied at a 
more fundamental level by interchange of the active and passive electric parts 
of the Riemann curvature$^{1}$. (Active and passive electric parts of the 
Riemann curvature are defined by the double (one for each 2-form) 
projection of the Riemann tensor and its double (both left and right) dual 
on a timelike unit vector, and dual is the usual Hodge dual, $*R_{abcd} = 
1/2 \epsilon_{abmn}R^{mn}_{cd}$). That is interchange of active ($E_{ab} = 
R_{acbd}u^cu^d$) and passive ($\tilde E_{ab} = *R*acbdu^cu^d$) electric 
parts implies interchange of the Ricci and Einstein 
tensors because contraction of Riemann gives Ricci while that of its 
double dual gives Einstein tensor. We have defined the electrogravity 
duality transformation$^{1}$ by interchange of the active and passive electric 
parts, $E_{ab}\leftrightarrow\tilde E_{ab}, H_{ab}\rightarrow H_{ab}$. Under 
this duality transformation it is clear that $\rho\leftrightarrow\rho_t, 
~\rho_n\rightarrow\rho_n, ~P_c\rightarrow P_c$.

 Then the condition $\rho_t = \rho_n = P_c =  0$ is electrograv-dual to 
the effective empty 
space equation given above, and its solution would give rise to the 
space dual to empty space. It can be easily verified that it integrates 
out to give the general 
solution given by $g_{00} = -g^{11} = 1 - 8\pi G \eta^2 - 2G M/r$, where 
$\eta$ is a constant. 
This is an asymptotically non-flat non-empty space which reduces to the 
Schwarzschild empty space for $\eta = 0$. At large $r$, the stresses it 
produces accord precisely to that of a global monopole of core 
mass $M$ and $\eta$ indicating the scale of symmetry breaking$^{2}$. 
Alternatively it can exactly for all $r$ represent a 
Schwarzschild black hole sitting in a cloud of string dust$^{3}$. It is 
remarkable that here it arises as dual to empty space,i.e dual to the 
Schwarzschild black hole$^{1}$. A global monopole 
is supposed to be produced when global symmetry $O(3)$ is spontaneously 
broken into $U(1)$ in phase transition in the early Universe. The 
physical properties of this space have been investigated$^{5}$ and it turns 
out that the basic character of the field remains almost the same except for 
scaling of the Schwarzschild's values for the black hole temperature, 
the light bending and the perihelion advance$^{6}$. The difference between 
the Schwarzschild solution and its dual can be demonstrated as 
follows. Both the solutions have $g_{00} = -g^{11} = 1 + 2\phi$ with 
$\bigtriangledown^2 \phi = 0$, which would have the general solution $\phi 
= k - M/r$. The Schwarzschild solution has $k = 0$, while the dual does 
not. This is the only essential difference between the two. It is this 
constant, which is physically trivial in the Newtonian theory, that 
brings in the global monopole charge, a topological defect.

 Let us also consider the remaining possibility, $\rho = \rho_t = P_c = 0$ 
which would in terms of the Ricci components imply $R=0, R^0_0 = 0$. This 
integrates out to give the general solution, $g_{00} = (k + \sqrt{1 - 
2GM/r})^2, 
g_{11} = - (1 - 2GM/r)^{-1}$, where $k$ is a constant. It is an 
asymptotically flat non-empty space with the stresses given by 

$$ T^1_1 =  {2kGM/r^3 \over k + \sqrt{1-2GM/r}} = - 2T^2_2 . $$

 Obviously, these stesses cannot correspond to any physically acceptable 
matter field because $\rho = 0$. On the other hand 
the spacetime unlike the dual solution remains asymptotically 
flat. It will admit a static surface only if $k < 0$ at $r_s = 
2GM/(1-k^2)$ and a horizon at $r_h = 2GM$. However $r \ge 2GM$ always 
for $g_{00}$ to be real. The region lying between $r_s$ and $r_h$ would 
define an ergosphere where negative energy orbits can, as for the Kerr black 
hole, occur. The Penrose process$^{7}$ can be set up to extract out the 
contribution of $k$ only if it is negative. However we do not know physical 
source for $k$. 

 On the other hand, when $k>0$, there occurs no horizon and it can 
represent a wormhole$^{8}$ of the throat radius $r = 2GM$. It is 
remarkable that that it has the basic character of a wormhole which 
needs to be further investigated. Pursuing on this track, we are presently 
working out a viable wormhole model$^{9}$.

 This space is certainly empty relative to timelike 
particles as both $\rho$ and $\rho_t$ vanish but not so for photons as 
$\rho_n \neq 0$. At the least, it can be viewed as an asymptotic 
flatness preserving perturbation to the Schwarzschild field. 

 Further it is also possible to characterize the 
Reissner-Nordstr${\ddot o}$m solution of a charged black hole by $\rho = 
\rho_t, \rho_n = P_c = 0$, and the de Sitter ($\Lambda$ - vacuum) space by 
$\rho + \rho_t = 0, ~\rho_n = P_c = 0$. In the Ricci components, the 
former would translate into $R = 0, R^0_0 = R^1_1$. This is clearly invariant 
under the duality transformation. It is a non-empty space with 
trace-free stress tensor. The de Sitter space is given by $\rho 
+ \rho_t = P_c = 0$, which implies $R_{ab} = \Lambda g_{ab}$. Of course 
under the duality transformation the sign of $\Lambda$ would change 
indicating that the de Sitter and anti de Sitter are dual of each-other.

 The next question is, could other empty space solutions representing 
isolated sources be characterized similarly? 

 It turns out that it is possible to characterize the NUT 
solution and its dual$^{10,11}$ in the similar manner. However an 
additional condition would come from the gravomagnetic monopole$^{12}$
character of the field. The most difficult and challenging problem 
would be to bring the Kerr solution in line. That is an open question and 
would engage us for some time in future. The crux of the matter is to 
identify the additional condition corresponding to gravomagnetic 
character of the field and solving the resulting equations. Once that is 
achieved, our new characterization of vacuum would cover all the 
interesting cases. 

 In conclusion we would like to say that it is always illuminating and 
insighful to understand the relativistic situations in terms of the 
familiar Newtonian concepts and constructs. Relating empty space to 
absence of energy and convergence density is undoubtedly physically 
very appealing and intuitively soothing. The most remarkable aspect of this 
way of looking at empty space is that it gives rise in a natural manner to 
the new spaces dual to the corresponding empty spaces. The dual spaces 
only differ from the orginal vacuum spaces by inclusion of a topological 
defect, global monopole charge.

 Note that the characterization of empty space and its dual is by  
the covariant equations. Earlier the dual spacetimes$^{1,13}$ were  
obtained by modifying the vacuum equation, so as to break the invariance 
relative to the electrogravity duality transformation, 
in a rather ad-hoc manner. 
Now the effective vacuum equation has the direct physical meaning in 
terms of the energy 
and convergence density. This characterization could as well be applied in  
lower and higher dimensions to find new dual spaces. For example, in 
3-dimensional gravity the dual space represents a new class of black hole 
spaces$^{14}$ with a string dust 
matter field. For higher dimensions, the method would simply go 
through without any change for n-dimensional spherically symmetric space 
and dual space would represent a corresponding Schwarzschild black hole 
with a global monopole charge. It can be further shown that a global 
monopole field in the Kaluza-Klein space can be constructed 
similarly$^{15}$ as dual to the vacuum solution$^{16}$. It is thus an 
interesting characterization of empty space which leads to new spaces
dual to corresponding empty spaces.

Acknowledgement: I thank the referee for his constructive comments.

\vfill\eject

\n{\bf References} 
\s 
\item{1.} Dadhich, N., Mod. Phys. Lett., 1999, {\bf A14}, 337. 
\item{2.} Barriola, M. and Vilenkin, A.,  Phys. Rev. Lett.,1989, {\bf 63}, 341. 
\item{3.} Letelier, P. S., Phys. Rev., 1979, {\bf D20}, 1294.
\item{4.} Raychaudhuri, A. K., Phys. Rev., 1955, {\bf 90}, 1123. 
\item{5.} Harari, D. and Lousto, C., Phys. Rev., 1990, {\bf D42}, 2626. 
\item{6.} Dadhich, N., Narayan, K. and Yajnik, U., Pramana, 1998, {\bf 50}, 
307. 
\item{7.} Penrose, R., Riv. Nuovo Cimento, 1969, {\bf 1}, 252. 
\item{8.} Visser, M., Lorentzian Wormholes: From Einstein To Hawking 
(American Institute of Physics, 1995).
\item{9.} Mukherjee, S. and Dadhich, N., under preparation. 
\item{10.} Dadhich, N. and Nouri-Zonoz, M., under preparation. 
\item{11.} Nouri-Zonoz, M, Dadhich, N. and Lynden-Bell, D., Class. 
Quant.  Grav., 1999, {\bf16}, 1021.
\item{12.} Lynden-Bell, D. and Nouri-Zonoz, M., Rev. Mod. Phys., 1998, {\bf 
70}, 427. 
\item{13.} Dadhich, N., in Black Holes, Gravitational Radiation and the 
Universe, eds. B. R. Iyer and B. Bhawal (Kluwer, 1999), p.171.
\item{14.} Bose, S., Dadhich, N. and Kar, S., A new class of black holes in 
2+1 - gravity, gr-qc/9911069, accepted in Phys. Lett. B.
\item{15.} Dadhich, N., Patel, L. K. and Tikekar, R., Global monopole as 
dual-vacuum solution in Kaluza-Klein spacetime, gr-qc/9909065, Mod. 
Phys. Lett., 1999, {\bf A14}, 2721.
\item{16.} Banerjee, A., Chatterjee, S. and Sen, A. A., Class. Quant. 
Grav., 1996, {\bf13}, 3141.

 \bye